\documentclass[12pt]{article}
\usepackage{epsf,rotating,latexsym}
\newcommand{\mc}[3]{\multicolumn{#1}{#2}{#3}}
\begin{document}

\begin{center}
{\Large \bf Three years of Galileo dust data: II. 1993 to 1995}
\end{center}


\bigskip

{\bf
        H.~Kr\"uger\footnote{{\em Correspondence to:} 
krueger@galileo.mpi-hd.mpg.de}, 
        E.~Gr\"un$^1$,
        D.~P.~Hamilton$^2$, 
        M.~Baguhl$^1$, 
        S.~Dermott$^3$,
        H.~Fech\-tig$^1$,
        B.~A.~Gustaf\-son$^3$,
        M.~S.~Hanner$^4$,
        M.~Hor\'anyi$^5$,
        J.~Kissel$^1$,
        B.~A.~Lind\-blad$^6$,
        D.~Linkert$^1$,
        G.~Linkert$^1$,
        I.~Mann$^7$,
        J.~A.~M.~McDonnell$^8$,
        G.~E.~Mor\-fill$^9$, 
        C.~Polanskey$^4$,
        R.~Riemann$^{10}$,
        G.~Schwehm$^{10}$,
        R.~Srama$^1$, \\
\centerline{
        and H.~A.~Zook$^{11}$
}}

\bigskip

\small
\begin{tabular}{ll}
1)& Max-Planck-Institut f\"ur Kernphysik, 69029 Heidelberg, Germany\\
2)& University of Maryland, College Park, MD\,20742-2421, USA\\
3)& University of Florida, Gainesville, FL\,32611, USA \\
4)& Jet Propulsion Laboratory, Pasadena, California 91109, USA\\
5)& Laboratory for Atmospheric and Space Physics, Univ.
                 of Colorado, Boulder, \\ 
  & CO\,80309, USA\\
6)& Lund Observatory, 221 Lund, Sweden\\
7)& Max-Planck-Institut f\"ur Aeronomie, 37191 Katlenburg-Lindau, Germany\\ 
8)& University of Kent, Canterbury CT2 7NR, UK\\
9)& Max-Planck-Institut f\"ur Extraterrestrische Physik, 85748 Garching, 
                                                                   Germany\\ 
10)& ESTEC, 2200 AG Noordwijk, The Netherlands\\
11)& NASA Johnson Space Center, Houston, Texas 77058, USA\\
\end{tabular}

\normalsize

\bigskip

\begin{abstract}

Between January 1993 and December 1995 the Galileo spacecraft 
traversed interplanetary space between Earth and Jupiter and 
arrived at Jupiter on 7 December 1995. The dust instrument 
onboard the spacecraft was operating during most of the 
time and data from the instrument were obtained via 
memory readouts which occurred at rates between twice per day and 
once per week. All events were classified by an onboard
program into 24 categories. Noise events were usually 
restricted to the lowest categories (class~0). During Galileo's 
passage through Jupiter's radiation belts on 7 December 1995 
several of the higher categories (classes~1 and 2) also show 
evidence for contamination by noise. The highest categories 
(class~3) were noise-free all the time. 
A relatively constant impact rate of interplanetary and 
interstellar (big) particles of 0.4 impacts per day was detected over 
the whole three-year time span. In the outer solar system (outside 
about 2.6~AU) they are mostly of interstellar origin, whereas 
in the inner solar system they are mostly interplanetary particles.
Within about 1.7~AU from Jupiter intense streams of small dust 
particles were detected with impact rates of up to 20,000 per day
whose  impact directions are compatible with a Jovian origin. 
Two different populations of dust particles were detected in
the Jovian magnetosphere: small stream particles during Galileo's
approach to the planet and big particles concentrated closer to 
Jupiter between the Galilean satellites. There is strong evidence 
that the dust stream particles are orders of magnitude smaller in 
mass and faster than the instrument's calibration, whereas the 
calibration is valid for the big particles. Because the data 
transmission 
rate was very low, the complete data set for only a small fraction 
(2525) of all detected particles could be transmitted to Earth; 
the other particles were only counted. Together with the 358 
particles published earlier, information about 2883 particles detected by 
the dust instrument during Galileo's six years' journey to Jupiter
is now available.

\end{abstract}

\section{Introduction}

The dust sensors onboard the Galileo and Ulysses spacecraft 
are highly sensitive impact ionization detectors. The two nearly 
identical sensors have been described in detail by Gr\"un et al. 
(1992a,b, 1995a). Results from the dust 
experiments on both spacecraft have 
been published frequently: Gr\"un et al. (1992c) published 
early results from both missions, and dust originating from comets and asteroids 
has been considered by Riemann and Gr\"un (1992), Hamilton and Burns 
(1992) and Gr\"un et al. (1994). Dust streams originating from the 
Jovian system and interstellar dust particles have been discovered 
with the Ulysses detector (Gr\"un et al. 1993). These 
were later confirmed by Galileo (Baguhl et al. 1995, Gr\"un et al. 
1996a). Gr\"un et al. (1996b) discuss dust particles detected a few 
days around Galileo's Io flyby on 7 December 1995. During its orbital 
tour within the Jovian magnetosphere Galileo has demonstrated the 
electromagnetic interaction of submicron-sized dust 
particles with Jupiter's magnetic field (Gr\"un et al. 1997, 1998). 
The data from both instruments -- Galileo and Ulysses -- have 
been used to model the interplanetary meteoroid populations 
(Divine 1993, Gr\"un et al. 1997c) and to compare the mass distribution 
of interstellar particles derived from in-situ measurements 
with that obtained from astronomical observations (Baguhl et al. 1996, 
Landgraf and Gr\"un 1998). Finally, Zook et al. (1996) and Hor\'anyi 
et al. (1997) used data from both spacecraft to model the Jovian dust 
streams. 
 
This is the fourth paper in a series dedicated to presenting both 
raw and reduced data from the Galileo and Ulysses dust instruments. 
Gr\"un et al. (1995a, hereafter Paper~I) describe the reduction process 
of Galileo and Ulysses data. Papers II and III (Gr\"un et al. 1995b,c) 
contain the data sets from the initial three and two years
of the Galileo and Ulysses missions, respectively. In the case of 
Galileo the time period covered (Paper~II) was December 1989 to 
December 1992. In the current paper we extend the Galileo data set 
from January 1993 until December 1995. In a companion paper 
(Kr\"uger et al. 1998, Paper~V) we publish the Ulysses data set for the 
same time period.
The main data products are a table of the impact rate of all impacts 
determined from the particle accumulators and a table of both raw and
reduced data of all ``big'' impacts received on the ground. The 
information presented in these papers is similar to data which we 
are submitting to the various data archiving centers (Planetary 
Data System, NSSDC, etc.). The only difference is that the paper version 
does not contain the full data set of the large number of ``small'' 
particles. Electronic access to the full data set is also possible via 
the world wide web: http://galileo.mpi-hd.mpg.de.

This paper is organised similarly to Paper~II. Section~\ref{mission} 
gives a brief overview of the Galileo mission until the end of 1995,
the dust instrument and lists important mission events during the 1993 
to 1995 period. A description of the Galileo dust data set for 1993 to 
1995 together with a discussion of the detected impact rate is given 
in Sect.~\ref{events}. Section~\ref{analysis} analyses and discusses 
various characteristics of the new data set. We dedicate Sect.~\ref{io} 
to an analysis of the dust particles and the noise events detected 
around Io flyby on 7 December 1995. 
Finally, Sect.~\ref{summary} summarizes our results.

\section{Mission and instrument operation} \label{mission}

Galileo was launched on 18 October 1989. During the initial 3 years of 
the mission Galileo was in the inner solar system and had flybys of 
Venus, Earth and the asteroid Gaspra. After its second Earth flyby in 
December 1992, Galileo had enough energy to leave the inner solar 
system, fly by the asteroid Ida, and reach Jupiter in December 1995. 
Galileo's interplanetary 
trajectory is shown in Fig.~\ref{trajectory} with a few important 
events indicated: on 28 August 1993 Galileo flew by the asteroid 
Ida; the atmospheric entry probe was released from the Galileo 
orbiter on 13 July 1995; on 7 December 1995 Galileo arrived at 
Jupiter and -- after a swing by at Io -- was injected into a highly 
elliptical orbit about the planet. Orbital elements of the Galileo
trajectory are provided in Tab.~\ref{elements}.
Galileo now performs regular close 
flybys of Jupiter's Galilean satellites. A detailed description of 
the Galileo mission and the spacecraft are given by Johnson et al. 
(1992) and D'Amario et al. (1992).

Galileo is a dual spinning spacecraft with an antenna that points 
antiparallel to the positive spin axis. During most of the initial 3 years 
of the mission the antenna pointed towards the Sun (cf. Fig.~2 in 
Paper~II). Since 1993 the antenna usually points towards Earth.
Deviations from the Earth pointing direction 
between January 1993 and December 1995 are shown in Fig.~\ref{pointing}.	

The Dust Detector System (DDS) is mounted on the spinning section of 
Galileo and the sensor axis is offset by an angle of $60^{\circ}$
from the positive spin axis (Fig.~\ref{galileo}, an angle of $55^{\circ}$
has been erroneously used earlier). The rotation angle 
measures the viewing 
direction of the dust sensor at the time of a dust particle impact. 
During one spin revolution of the spacecraft the rotation angle scans 
through a complete circle of $360^{\circ}$. At rotation angles of 
$90^{\circ}$ and $270^{\circ}$ the sensor axis lies nearly 
in the ecliptic plane, and at $0^{\circ}$ it is close to ecliptic north. 
DDS rotation angles are taken positive around the negative spin axis of 
the spacecraft. This is done to easily compare Galileo impact spin 
angle data with those taken by Ulysses, which, unlike
Galileo, has its positive spin axis pointed towards Earth.
DDS has a $140^{\circ}$ wide field of view and during one spin revolution
of the spacecraft the sensor axis scans a cone with $120^{\circ}$
opening angle towards the anti-Earth direction. Dust particles that arrive
from within $10^{\circ}$ of the positive spin axis (anti-Earth direction) 
can be sensed at all rotation angles, whereas those that arrive at angles 
from $10^{\circ}$ to $130^{\circ}$ from the positive spin axis can only 
be detected over a limited range of rotation angles.

During most of the interplanetary cruise (i.\,e. prior to 7 December 
1995) we received DDS data as instrument memory-readouts (MROs).
The MROs returned event data which had accumulated over time in the 
instrument memory. Initially, an MRO contained 14 instrument 
data frames (with each frame comprising the complete data set of an 
impact or noise event, consisting of 128 bits, plus ancillary and 
engineering data). Since June 1990, when DDS was reprogrammed for 
the first time after launch, 
an MRO contains 46 instrument data frames (cf.~Paper~I). DDS time-tags
each impact event with an 8 bit word allowing for the identification of 
256 unique times. The step size of this time word was also changed 
from 1.1~h to 4.3~h in the June 1990 reprogramming to allow for longer 
time periods between MROs without loss of the impact time information. 
The total accumulation time is now $\rm 256 \times 4.3\,h = 46$days after 
which the time word is 
reset and the time labels of older impact events become ambiguous. 
MROs usually occurred twice a week which was sufficient to get the 
time information of the impact events transmitted to Earth within 
the 46 day period. The accuracy, however, with which the impact time can be 
determined, is limited to 4.3~h. 

For two periods of a few hours around Io flyby on 7 December 1995 the 
instrument was read-out every few minutes, the data were stored on Galileo's 
tape recorder and transmitted to Earth during the following months 
(record mode). 7 different instrument data frames were read-out this way 
within about one minute (with 6 frames containing the information of the 6 
most recent impact events, the 
so-called A range, cf.~Paper~I). Although fewer data frames were read-out 
in this manner at 
a time, the number of new events that could be transmitted to Earth 
in a given time period was much larger than with MROs due to the higher 
read-out cycle. 
Furthermore, in record mode the read-out cycle determines the accuracy 
of the impact time to within a few minutes, much better than with MROs.

Table~\ref{event_table} lists significant mission and dust instrument 
events for the period 1993 to 1995. A comprehensive list of earlier 
events can be found in Paper~II. After Galileo's second Earth flyby on 8 
December 1992, DDS was 
brought into its nominal operational mode for the rest of the 
interplanetary cruise to Jupiter: the channeltron voltage was set to 
1020~V (HV~=~2), 
the event definition status was set such that the channeltron or 
the ion-collector channel can independently initiate a measurement cycle 
(EVD~=~C,I) and the detection thresholds for ion-collector, channeltron, 
electron-channel and entrance grid were set (SSEN~=~0,~0,~1,~1). 
Detailed descriptions of these symbols are given in Paper~I. 

The operational mode of DDS was changed several times during noise          
tests between 1993 and 1995: 
starting from the nominal configuration described above, all tests have been 
achieved with the same instrument settings. The following changes 
of the instrument configuration were applied at 2 to 3-day 
intervals:
a) increase the channeltron high voltage by one digital step (HV~=~3);
b) reset the channeltron high voltage to its nominal value (HV~=~2); 
c) set the event definition status such that the channelton, the ion 
collector and the electron-channel can each, independently, initiate a 
measurement cycle (EVD~=~C,I,E);
d) set the thresholds for all channels to their lowest levels 
(SSEN~=~0,~0,~0,~0);
e) reset the thresholds and the event definition status to their nominal 
configuration (SSEN~=~0,~0,~1,~1, EVD~=~C,I).
Note that after step e) DDS is brought back to its nominal configuration. 
No detectable changes in the noise characteristics were
revealed by these noise tests.

The instrument memory was not read out between 3 and 28 July 1995 and 
no DDS data could be obtained. In this period 
the atmospheric entry probe was released from the orbiter and a 
propulsion system burn occurred during the orbit deflection maneuver (ODM)
of the orbiter. Within the 
Jovian magnetosphere a strong increase in the high energy electron flux 
was expected close to Jupiter. To save the instrument from the 
hazards associated with Jupiter's radiation environment the 
channeltron voltage was reduced and the detection thresholds were 
increased on 6 December 1995, 5:40h (HV\,=\,1, EVD\,=\,I, 
SSEN\,=\,2,\,0,\,2,\,2) at a distance of $\rm 30\,R_J$ from Jupiter 
(Jupiter radius, $\rm R_J = 71,492~km$). On 7 December 1995, 23:25\,h, 
shortly before insertion of Galileo into Jupiter orbit,  the channeltron 
high voltage was switched off. 
The Io and Jupiter flybys will be discussed in detail in Sect.~\ref{io}. 

\section{Impact events and classification scheme} \label{events}

DDS classifies all events -- real impacts of dust particles and noise 
events -- into one of 24 different categories (6 amplitude ranges for
the charge measured on the ion collector grid 
and 4 event classes) and counts them in 24 corresponding accumulators 
(Paper~I). Most of the 24 categories are relatively free from noise 
and only sensitive to real dust impacts, except for extreme situations 
like the crossings of the radiation belts of Earth, Venus (Paper~II) and 
Jupiter (7 December 1995, Sect.~\ref{io}). During most of Galileo's 
initial three years of interplanetary cruise since launch only the lowest 
amplitude and class categories -- AC01 (event class 0, amplitude 
range 1, AR1), AC11, and AC02 -- were contaminated by noise events 
(Paper~II). 

In a detailed analysis of the Ulysses data set published in Paper~III, 
Baguhl et al. (1993) identified a large number of ``small'' impacts in 
the three lowest categories. They deduced a modified event 
classification scheme that allows for a better discrimination between 
noise events and real dust impacts. The modified 
scheme was loaded to the instrument computer on board Galileo during the second 
reprogramming of DDS on 14 July 1994 and is shown in Tab.~\ref{class_scheme}. 
The definition of class 3 remains unchanged with respect to the 
old scheme published in Paper~I. Classes~1 and 2 are now divided into two 
subclasses. With the modified scheme noise events 
are now usually restricted to class 0. Class 3 always contains good dust 
impacts (two AC31 events were rejected which occurred during a noise 
test on 15 August 1995 because they did not fulfill the class~3 classification 
criteria). After the 14 July 1994 reprogramming, all class~1 and 2 events 
detected in the low radiation environment of interplanetary space are 
true dust impacts.
Dust impacts which do not fulfill the criteria of classes 1, 2 or 3 are 
automatically assigned class~0. Therefore, class 0 may still contain good
dust impacts, especially in the higher amplitude ranges. Although noise 
events are now normally restricted to class 0, classes 1 and 2 are also 
contaminated by noise in extreme radiation environments (Sect.~\ref{io}). 
With the modified scheme the mass sensitivity of 
the instrument could be improved by a factor of eight and the number of 
dust particles identified in the Ulysses data set from October 1990 
to December 1992 was enhanced from 333 to 968 (Baguhl et al. 1993). 

Table~4 lists the number of all dust impacts identified with the Galileo 
dust sensor between 1 January 1993 and 31 December 1995. Before 14 July 1994
the particles were classified with the old classification scheme 
whereas afterwards the modified scheme was applied. The number of 
impacts is typically given in intervals of about one week, depending
on the occurrence of MROs. When the frequency of MROs was higher or 
when no impact was recorded, MROs were put together. Typically, MROs
occurred twice per week. In interplanetary space where the
impact rate was roughly one impact per two days (see below)
this was sufficient to receive the complete information of all particles.
During the occurrence of dust streams when the impact rates 
in the lowest amplitude range (AR1) were higher by several
orders of magnitude the full information of only a small fraction of 
all detected particles could be transmitted to Earth. In this case 
the impact rates had to be deduced from the accumulators. 

The frequency of MROs limits the maximum impact rate of dust particles 
that can be determined from the accumulators. If an unknown number 
of accumulator overflows occurs between individual MROs the number of particles and, 
hence, the impact rate deduced is only a lower limit to the real dust
impact rate. With MROs occurring every few days, impact rates of up to
100 per day could be determined from the accumulators. 
Between the end of July and October 
1995, when the strongest dust streams were observed, MROs were 
transmitted to Earth usually daily, sometimes even more frequently. 
During the occurrence of the most intensive dust streams MROs were 
split-up and transmitted in two segments separated by about 25~min. 
This way rates of up to 20,000 impacts 
per day could be determined from the accumulators over such short time 
intervals. Entries in Tab.~4 indicated by ''{\footnotesize \#}'' signs give 
the number of impacts determined from the accumulators in the lowest 
amplitude range over the 25~min interval. No overflows of the 
accumulators for the higher amplitude ranges (AR2 to AR6) 
occurred between MROs even during the most intensive dust streams.
During the strongest dust streams, however, the effective measurement 
time for such particles was significantly reduced due to deadtime (cf. 
Fig~\ref{rot_angle}). 

In this paper the terms ''small`` and ''big`` do {\em not} have the
same meaning as in Paper~II. Here we call all particles in classes 1,
2 and 3 in the amplitude ranges 2 and higher (AR2 to AR6) ''big``. Particles in
the lowest amplitude range (AR1) are called ''small``. This distinction
separates the small Jovian dust stream particles from big particles of
interplanetary or interstellar origin (cf. Fig.~\ref{rot_angle}).

The total dust impact rate recorded by DDS from 1993 to 1995 is shown in 
Fig.~\ref{rate}. During this three year time span the average impact 
rate of big particles (AR2 to AR6) was rather constant with 0.4 impacts 
per day. From the beginning of 1993 until the first half of 1994 the 
average impact rate of small particles (AR1) was about an order of 
magnitude lower. Later the small particles dominated the overall impact
rate. No increase in the impact rate was detected 
during the Ida flyby on 28 August 1993 and during the passage through 
the asteroid belt. This is consistent with the absence of an enhanced impact 
rate during the Gaspra flyby two years earlier, and it is in agreement with the 
predictions of Hamilton and Burns (1992). DDS detected the 
first Jovian dust stream on 25 June 1994 with a peak rate of 
10 impacts per day. At this time Galileo was still about $\rm 1.7~AU$
away from Jupiter. During the dust streams detected later and closer
to Jupiter an impact rate of up to 20,000 per day has been detected. 
Although such an impact rate is close to the technical limit of Galileo, the 
data indicate that undetected accumulator overflows did not occur 
frequently. A detailed discussion of the Jovian dust 
streams detected with DDS is given by Gr\"un et al. (1996a). 

Table~5 lists all 395 big particles detected in classes 1 to 3 between January 
1993 and December 1995 for which the complete information exists (Note that 
this table includes 47 class~1 and 2 events around Io and Jupiter flybys 
which are possibly noise events, see Sect.~\ref{io}). 
We do not list the small particles (AR1) in Tab.~5 because their masses and 
velocities are outside the calibrated range of DDS. The stream particles are 
believed to be about 10~nm in size and their velocities exceed 200~km/s 
(Zook et al. 1996).
Any mass and velocity calibration of these particles 
would be unreliable. The complete information of a total of 2130 small 
particles has been transmitted to Earth from 1993 to 1995. The full 
data set of all 2525 small and big particles is available in electronic 
form.

In Tab.~5 dust particles are identified by their sequence number 
and their impact time. Gaps in the sequence number are due to the 
omission of the small particles. The event category -- class (CLN) 
and amplitude range (AR) -- are given. Raw data as transmitted to Earth 
are displayed in the next columns: sector value (SEC) which is the
spacecraft spin orientation at the time of impact, 
impact charge numbers (IA, EA, CA) and rise times (IT, ET), time
difference and coincidence of electron and ion signals (EIT, EIC),
coincidence of ion and channeltron signal (IIC), charge reading at
the entrance grid (PA) and time (PET) between this signal and
the impact. Then the instrument configuration is given: event
definition (EVD), charge sensing thresholds (ICP, ECP, CCP, PCP) and
channeltron high voltage step (HV). See Paper~I for further
explanation of the instrument parameters. 

The next four columns in Tab.~5 give information about Galileo's orbit: 
heliocentric distance (R), ecliptic longitude and latitude (LON, LAT) 
and distance from Jupiter ($\rm D_{Jup}$). The next column gives the 
rotation angle (ROT) as described in Sect.~\ref{mission}. 
Whenever this value is unknown, ROT is arbitrarily set to
999. This occurs 21 times (80 times in the full data set that includes 
the small particles). Then follows the pointing direction of DDS at 
the time of particle impact in ecliptic longitude and latitude 
($\rm S_{LON}$, $\rm S_{LAT}$).
When ROT is not valid $\rm S_{LON}$ and $\rm S_{LAT}$ are also useless. 
Mean
impact velocity (V) and velocity error factor (VEF, i.e. multiply or
divide stated velocity by VEF to obtain upper or lower limits) as well as mean 
particle mass (M) and mass error factor (MEF) are given in the last 
columns. For VEF $> 6$, both velocity and mass values should be
discarded. This occurs for 5 impacts. No intrinsic dust charge values 
are given (see Svestka et al. 1996 for a detailed analysis).

\section{Analysis} \label{analysis}

The positive charge measured on the ion collector, $\rm Q_I$, is 
the most important impact parameter determined by DDS because it is 
rather insensitive to noise. Figure~\ref{nqi} shows the distribution of 
$\rm Q_I$ for the full data set (small and big particles) from 1993 to 
1995. Ion impact charges have been detected over the entire range of six 
orders of magnitude that the instrument can measure. Two impacts (about 
0.1\% of the total) are close to the saturation limit of $\rm Q_I \sim 
10^{-8}\,C$ and may thus 
constitute lower limits of the actual impact charges. The impact charge 
distribution of the big particles ($\rm Q_I > 10^{-13}\,C$) follows a 
power law with index -0.43 and is shown as a dashed line. This slope
is steeper than the value of -1/3 given for Galileo in Paper~II and 
flatter than the -1/2 given for Ulysses in Paper~III. It indicates  
that, on average, Galileo has detected smaller particles 
in the outer solar system than in the inner solar system. This is in 
agreement with a larger 
contribution of interstellar particles further away from the Sun.
Note that the Jovian stream particles (AR1) have been excluded from the
power law fit. 

In Fig.~\ref{nqi} the small particles ($\rm Q_I < 10^{-13}\,C$) 
are put together in two histogram bins. To analyse their behavior 
in more detail, their number per individual digital step 
is shown separately in Fig.~\ref{nqi2}. The distribution flattens 
for impact charges below $\rm 2\times 10^{-14}\,C$. This 
indicates that the sensitivity threshold of DDS may not be sharp and the 
number of impacts with the lowest impact charges $\rm Q_I$ 
may not be complete. The impact charge distribution for these small particles 
above $\rm Q_I > 2\times 10^{-14}\,C$ follows a power law with index 
-1.9. This indicates that the size distribution of the small stream 
particles rises steeply towards smaller particles and is much 
steeper than the distribution of the big particles shown in Fig.~\ref{nqi}. 

The ratio of the channeltron charge $\rm Q_C$ and the ion collector
charge $\rm Q_I$ is a measure of the channeltron amplification A which
is an important parameter for the dust impact identification (Paper~I).
The in-flight channeltron amplification was determined 
in Paper~II for the initial three years of the 
Galileo mission. For a channeltron high voltage of 1020~V (HV~=~2) 
the amplification $\rm Q_C/Q_I$ obtained for $\rm 
10^{-12}{\rm\, C} \le Q_I \le 10^{-10}{\rm\, C}$ was $\rm A \sim 1.6$. 
Here we repeat the same analysis for the time period 1993 to 1995 
to identify any degrading of the channeltron. Figure~\ref{qiqc}
shows the charge ratio $\rm Q_C/Q_I$ as a function of 
$\rm Q_I$ for the same high voltage as in Paper~II. The charge 
ratio $\rm Q_C/Q_I$ determined for $\rm 10^{-12}{\rm\,C} \le Q_I 
\le 10^{-10}{\rm\,C}$ is $\rm A \sim 1.4$. Thus, no significant aging 
of the channeltron is detectable. We neglect the large 
number of small particles in the lowest amplitude range in 
Fig.~\ref{qiqc} because they do not contribute to the determination 
of the channeltron amplification. Their neglect better illustrates 
the number distribution of impacts in the higher amplitude ranges. 

Figure~\ref{mass_speed} displays the masses and velocities of 
all dust particles detected between 1993 and 1995. As in the 
earlier period (1990 to 1992) velocities occur over 
the entire calibrated range from 2 to 70 km/s and the masses 
vary over 10 orders of magnitude from $\rm 10^{-6}$ to 
$\rm 10^{-16}\,g$. The mean errors are a factor of 2 for the 
velocity and a factor of 10 for the mass. The clustering
of the velocity values is due to discrete steps in the rise
time measurement but this quantization is much smaller than the
velocity uncertainty. Masses and velocities in the lowest 
amplitude range (particles indicated by plus signs) should be 
treated with caution. These are mostly Jovian stream 
particles for which we have clear indications that their masses 
and velocities are outside the calibrated range of DDS (Zook 
et al. 1996). The particles are probably much faster and smaller 
than implied by Fig.~\ref{mass_speed}. On the other hand, the 
mass and velocity calibration is valid for the bigger particles. 
For many particles in the lowest two amplitude ranges (AR1 and 
AR2) the velocity had to be computed from the ion charge signal 
alone which leads to the striping in the lower mass range in 
Fig.~\ref{mass_speed} (most 
prominent above 10 km/s). In the 
higher amplitude ranges the velocity could normally be calculated 
from both the target and the ion charge signal which leads to  
a more continuous distribution in the mass-velocity plane. Impact 
velocities below about 3 km/s should be treated with caution
because anomalous impacts onto the sensor grids or structures 
other than the target generally lead to prolonged rise times and 
hence to artificially low impact velocities. 

The sensor orientation (rotation angle) at the time of particle 
impact is shown in Fig.~\ref{rot_angle}. Particles approaching 
parallel to the ecliptic plane are detected at rotation 
angles of $90^{\circ}$ and $270^{\circ}$. Contour 
lines illustrate the detector sensitivity for interstellar particles. 
The impact direction of most of the big particles (filled circles) is 
compatible with the interstellar direction. The remaining big 
particles are compatible with an interplanetary origin.  
Baguhl et al. (1996) showed that in the outer solar system (i.~e. 
outside about 2.6~AU from the Sun) interstellar particles can be 
clearly distinguished from particles having interplanetary origin. 
At distances between 1 and 2.6~AU, however, interplanetary dust 
particles on prograde orbits approach from the same direction as 
interstellar particles and both populations cannot be separated by
impact direction arguments. In contrast to the small stream 
particles which are outside the calibrated range of DDS,
the calibration is valid for the interplanetary and interstellar 
particles (cf. Tab.~5).

The small particles (plus signs) cluster at rotation angles 
between $200^{\circ}$ and $340^{\circ}$ which is compatible with 
a Jovian origin (Gr\"un et al. 1996a). The striping of small impacts 
(plus signs) is due to the occurrence of individual dust streams and 
a comparison with Fig.~\ref{rate} shows that the times of the 
stripes are coincident with the times of high impact rates. At the 
end of 1995 the impact rates were so high over several weeks that 
the symbols form a black area in Fig.\ref{rot_angle}.

\section{Io and Jupiter flybys} \label{io}

On 7 December 1995 Galileo arrived at Jupiter after its six years 
journey through interplanetary space. Because of
the expected increase in the high energy electron flux near Jupiter
the channeltron high voltage of DDS was reduced and the detection 
thresholds were increased on 6 December, 5:40~h (Tab.~\ref{event_table}).
At that time Galileo was at a distance of $\rm 15\,R_J$ from Jupiter
(Jupiter radius, $\rm R_J = 71,492~km$). The change in the instrument 
configuration let to a reduction in sensitivity by about a factor of six. 
On 7 December, 17:46~h Galileo flew by Io and at 21:54~h nearest to Jupiter.
At 23:25~h the channeltron high voltage was switched off
and at 00:27~h on 8 December the orbit insertion maneuver was started
which brought Galileo into a bound orbit about Jupiter.

From 1 to 6 December 1995 MROs occurred about once per day. 
Around the time of Galileo's closest approach to Io on 7 December 
DDS data were read-out every few minutes and recorded to Galileo's tape 
recorder (record period, 15:21~h to 18:25~h). After Jupiter
closest approach (21:54~h) data were recorded for another 
two hours (23:22~h to 01:26~h). The recorded data from both flybys were 
transmitted to Earth a few month later. Initial results from the Io and 
Jupiter flybys have been published by Gr\"un et al. (1996b). 

Figure~\ref{rate_io} shows the dust impact rate before and around the 
Io and Jupiter flybys. During Galileo's approach to Jupiter 
the impact rate increased considerably and reached a maximum of about 
200 impacts per day on 4 December 1995 (day 338). Two days later the 
sensitivity of the instrument was reduced and the impact rate dropped 
by about a factor of 10. 
After the sensitivity reduction a few small particles were still 
being detected until closest approach to Io. At Io closest approach 
small particle impacts ceased. Big particles were only detected 
a few days before closest approach to Io when Galileo was 
within about $\rm 25\,R_J$ of Jupiter. The big particles 
show a concentration towards the inner Jovian system. 
Impacts of big particles were seen after Io flyby until closest 
approach to Jupiter, i.e. when impacts of small class~3 particles had 
already terminated. High impact rates of small particles and increases in 
the impact rate upon approach towards the inner Jovian system were also seen 
later during Galileo's orbital tour about the planet (Gr\"un et al. 
1997, 1998). Concentrations of big particles between 
the Galilean satellites have also been found. For the calculation of 
the impact rate in Fig.~\ref{rate_io} we have only considered 
the class~3 impacts because classes~1 and 2 show evidence for 
contamination by noise in the inner Jovian system (see below). 
The inclusion of big class~2 events -- which seem to be less affected 
by noise -- would increase the number of impacts per given time interval 
(cf. Tab.~3) but would not change our conclusions about the cessation of 
small dust impacts after Io and Jupiter flybys (Gr\"un et al. 1996b). 

In Fig.~\ref{rot_angle_io_1} we show the rotation angle of the 
class~3 impacts for which the complete information has been transmitted
to Earth. When Galileo was approaching Jupiter the 
impact directions of dust particles were concentrated between 
$\rm 210^{\circ}$ and $\rm 350^{\circ}$. About one day before 
closest approach to Io two particles also arrived from 
the opposite direction. The striping before day 341 is due to the 
occurrence of MROs once per day -- which allow for a time 
resolution of 4.3~h -- and the fact that the instrument 
memory of DDS can store only 16 class~3 events. Note that each vertical 
band of particles between days 332 and 340 in Fig.~\ref{rot_angle_io_1} 
corresponds to one discrete MRO. If there are 
more than 16 impacts between two MROs, the oldest events are lost. 
Before day 341, 15:20~h many class~3 particles have probably been lost. 
Note that the particle on day 339.0 that caused the peak 
in the impact rate of big particles in Fig.~\ref{rate_io} cannot 
be shown because its full information has not been transmitted to 
Earth.

So far we have only considered class~3 impacts around Io and Jupiter flybys. 
During closest approach to Jupiter in December 1995 and during all
later orbits high noise rates occurred when Galileo was within about 
$\rm 20~R_J$ of Jupiter (cf. Gr\"un et al. 1997). In
Fig.~\ref{rot_angle_io_2} we show the rotation angle for all impact events 
in classes~1 to 3 for which the complete information is available  
for a period of half a day around closest approach to Io.
The striping at 15:20~h and 20:40~h is again due to discrete instrument 
readouts and the time resolution is usually 4.3~h. Because of the switch to 
record mode which occured twice on day 341 the impact time can be determined with 
a higher accuracy from the internal timer of DDS: particles with impact times 
between 15:10~h and 15:20~h must have impacted the detector between 14:11~h and 
15:21~h. Their uncertainty in impact time is only 70~min. Particles with impact 
times in the time interval 18:30~h to 22:49~h have the full 4.3~h uncertainty, 
and those with impact times between 23:10~h and 23:22~h must have been 
detected between 22:49~h and 23:22~h, thus their timing uncertainty is only 
33~min.

The class~1 impact events detected about 3~hours before and at Io 
closest approach itself are spread over the whole range in rotation 
angles. At the same time the noise counter for the electron channel,
which counts all threshold exceedings on that channel, detected 
a high noise rate in excess of 300 noise events per second. 
Furthermore, the class~1 event rate is strongly peaked at Io 
closest approach (Fig.~\ref{rate_io_2}). The event rate at Jupiter 
closest approach cannot be shown because no high resolution 
recorded data were obtained for that period. 

The spread in rotation angle, the increased noise in the electron channel 
and the peak in impact rate 
indicate that at least some, if not all, of these events 
are due to noise rather than dust particle impacts. Although DDS 
can detect dust particles approaching from within $\rm 10^{\circ}$
of Galileo's positive spin axis at all rotation angles resulting in 
a $\rm 360^{\circ}$ spread in a rotation angle plot, one can 
hardly imagine a population of dust particles that caused only 
class~1 events and none in class~3. If the class~1 events are due 
to real dust particles approaching from close to the positive spin 
direction, one should also see a similar spread in rotation angle 
for the class~3 particles 
which is not the case. The smallest class~2 events (AR1) seem to 
follow a similar behavior as the class~1 events although much less 
obvious. Class~2 events also peak at Io closest approach but less 
strongly which is consistent with class~2 being less contaminated by 
noise than class~1.

Similar signatures of noise contamination within about $\rm 20\,R_J$ 
from Jupiter are evident in the DDS data from Galileo's later orbital 
tour in 
the Jovian system (cf. Gr\"un et al. 1997). Therefore, all class~1 and 
2 events detected in the Jovian system should be treated 
with caution. This applies to 26 class~1 and 20 class~2 events
in Tab.~5 which were detected on days 341 to 343, and to 50 events 
in each class if one includes AR1 events and which are published 
only electronically.
Only class~3 events can be considered good dust impacts at 
the moment. A detailed analysis of the noise characteristics of the 
DDS data from within the Jovian system is forthcoming.

\section{Summary} \label{summary}

In this paper, which is the fourth in a series of Galileo and 
Ulysses papers, we present data from the Galileo dust instrument for the 
period January 1993 to December 1995 when Galileo was traversing 
interplanetary space between Earth and Jupiter. The complete information 
(i.~e. all impact parameters measured by the dust instrument) for 
395 big particles detected during this period 
is given (including 47 class~1 and 2 events around Io and Jupiter 
flybys which are possibly noise events). The full data set that contains 
2525 small and big particles 
is available in electronic form only. Together with the 358 particles
published in Paper~II a set of 2883 particles detected during 
Galileo's six years' journey to Jupiter is now available. Our results can be 
summarized as follows: 

1) A relatively constant impact rate of interplanetary and 
interstellar particles of 0.4 impacts per day was detected over 
the whole three-year time span. In the outer solar system (outside 
about 2.6~AU) the big particles are mostly of interstellar origin, whereas 
in the inner solar system interplanetary particles dominate. These 
particles are in the calibrated mass and velocity range of DDS.

2) No increase in impact rate could be detected during the flyby
at the asteroid Ida confirming earlier results from the 
Gaspra flyby. 

3) Within about 1.7~AU from Jupiter intense streams of small dust 
particles were detected with impact rates of up to 20,000 per day whose  
impact directions are compatible with a Jovian origin. There is strong 
evidence that the dust stream particles are orders of magnitude smaller in 
mass and faster than the instrument's calibration, whereas the
calibration can be safely applied to the bigger interstellar and 
interplanetary particles.

4) Two different populations of dust particles were detected in
the Jovian magnetosphere: a) small stream particles during Galileo's
approach to the planet with impact rates of up to 200 per day 
3 days before Io flyby and b) bigger particles concentrated closer to 
Jupiter between the Galilean satellites. 

5) The data from the dust instrument obtained within about 
$\rm 20\,R_J$ from Jupiter (Jupiter radius, $\rm R_J = 71,492~km$) show 
evidence for contamination by noise. The behavior of the impact 
directions of class~1 and class~2 events differs from that of 
class~3 events. Class~1 and 2 events from the Jovian system should 
therefore be treated with caution. Class~3 events do not show any 
indication for noise contamination and are considered good dust 
impacts.

6) Noise tests performed regularly during the 3 years period did not 
reveal any change in the instrument noise characteristics. No 
degrading of the channeltron was revealed.

\hspace{1cm}

{\bf Acknowledgments.}
We thank the Galileo project at JPL for effective and successful 
mission operations. This work has been supported by the Deutsche 
Agentur f\"ur Raumfahrtangelegenheiten (DARA).

\section*{References}

{\small

{\bf Baguhl, M., Gr\"un, E., Linkert, D., Linkert, G.\ and Siddique, N.,}
Identification of 'small' dust impacts in the Ulysses dust
detector data. {\em  Planet.\ Space Sci. }{\bf 41}, No. 11/12, 1085-1098, 1993

{\bf Baguhl, M., Gr\"un, E., Hamilton, D.P., Linkert, G., Riemann, R.,
Staubach, P.\ and Zook H.,} The flux of interstellar dust observed by
Ulysses and Galileo. {\em Space Sci.\ Rev. }{\bf 72}, 471-476,1995

{\bf Baguhl, M., Gr\"un, E., Landgraf, M.} In situ measurements of 
interstellar dust with the Ulysses and Galileo spaceprobes. 
{\em Space Sci Rev.}, {\bf 78}, 165-172, 1996

{\bf D'Amario, L.A., Bright, L.E.\ and Wolf, A.A.,} Galileo trajectory
design. {\em Space Sci.\ Rev. }{\bf 60}, 23-78, 1992

{\bf Divine, N.\,} Five populations of interplanetary meteoroids.
{\em J.~Geophys.~Res. }{\bf 98}, 17029-17048, 1993



{\bf Gr\"un, E., Fechtig, H., Hanner, M.S., Kissel, J., Lindblad, B-A.,
Linkert, D., Linkert, G., Morfill, G.E.\ and Zook, H.A.,}
The Galileo Dust Detector. {\em Space Sci.\ Rev. }{\bf 60}, 317-340, 1992a

{\bf Gr\"un, E., Fechtig, H., Giese, R.H., Kissel, J., Linkert, D.,
Maas, D., McDonnell, J.A.M., Morfill, G.E., Schwehm, G.\ and
Zook, H.A.,} The Ulysses dust experiment.
{\em Astron.\ Astrophys.\ Suppl.\ Ser. }{\bf 92}, 411-423, 1992b

{\bf Gr\"un, E., Baguhl, M., Fechtig, H., Hanner, M.S., Kissel, J.,
Lindblad, B.-A., Linkert, D., Linkert, G., Mann, I., McDonnell, J.A.M.,
Morfill, G.E., Polanskey, C., Riemann, R., Schwehm, G., Siddique, N. and
Zook, H.A.,} Galileo and Ulysses dust measurements: From Venus
to Jupiter. {\em Geophys.\ Res.\ Letters }{\bf 19}, 1311-1314, 1992c

{\bf Gr\"un, E., Zook, H.A., Baguhl, M., Balogh, A., Bame, S.J.,
Fechtig, H., Forsyth, R., Hanner, M.S., Horanyi, M., Kissel, J.,
Lindblad, B.-A., Linkert, D., Linkert, G., Mann, I., McDonnell, J.A.M.,
Morfill, G.E., Phillips, J.L., Polanskey, C., Schwehm, G., Siddique, N.,
Staubach, P., Svestka, J. and Taylor, A., } Discovery of jovian
dust streams and interstellar grains by the Ulysses spacecraft. {\em
Nature }{\bf 362}, 428-430, 1993

{\bf Gr\"un, E., Hamilton, D.P., Baguhl, M., Riemann, R., Horanyi, M.\ and
Polan\-skey, C.,} Dust streams from comet Shoemaker-Levy 9? {\em
Geophys.\ Res.\ Let. }{\bf 21}, 1035-1038, 1994

{\bf Gr\"un, E., Baguhl, M., Fechtig, H., Hamilton, D.P., Kissel, J.,
Linkert, D., Linkert, G.\ and Riemann, R.,}
Reduction of Galileo and Ulysses dust data.
{\em Planet. Space Sci.} {\bf 43}, 941-951, 1995a (Paper I)

{\bf Gr\"un, E., Baguhl, M., Divine, N., Fechtig, H., Hamilton, D. P.,
Hanner, M. S., Kissel, J., Lindblad, B.-A., Linkert, D., Linkert, G., 
Mann, I., McDonnell, J. A. M., Morfill, G. E., Polanskey, C., Riemann, R.,
Schwehm, G., Siddique, N., Staubach P. and Zook, H. A.}
Three years of Galileo dust data. {\em Planet. Space Sci.} {\bf 43}, 953-969,
1995b (Paper II)

{\bf Gr\"un, E., Baguhl, M., Divine, N., Fechtig, H., Hamilton, D.P.,
Hanner, M.S., Kissel, J., Lindblad, B.-A., Linkert, D., Linkert, G.,
Mann, I., McDonnell, J.A.M., Morfill, G.E., Polanskey, C.,
Riemann, R., Schwehm, G., Siddique, N., Staubach, P.\ and Zook,
H.A., } Two years of Ulysses dust data. {\em Planet. Space Sci.} 
{\bf 43}, 971-999, 1995c (Paper III)

{\bf Gr\"un, E., Baguhl, M., Hamilton, D. P., Riemann, R., Zook, H. A.,
Dermott, S., Fechtig, H., Gustafson, B. A., Hanner, M. S., Horanyi, M.,
Khurana, K. K., Kissel, J., Kivelson, M., Lindblad, B.-A., Linkert, D., 
Linkert, G., Mann, I., McDonnell, J. A. M., Morfill, G. E., Polanskey, 
C., Schwehm, G. and Srama, R.} Constraints from Galileo observations on 
the origin of jovian dust streams. {\em Nature} {\bf 381}, 395-398, 
1996a

{\bf Gr\"un, E., Hamilton, D. P., Riemann, R., Dermott, S., Fechtig, H., 
Gustafson, B. A., Hanner, M. S., Heck, A., Hor\'anyi, M., Kissel, J.,
Kr\"uger, H., Lindblad, B.-A., Linkert, D., Linkert, G., Mann, I.,
McDonnell, J. A. M., Morfill, G. E., Polanskey, C., Schwehm, G., 
Srama, R. and Zook, H. A.} Dust measurements during Galileo's approach
to Jupiter and Io encounter. {\em Science} {\bf 274}, 399-401, 1996b.

{\bf Gr\"un, E., Kr\"uger, H., Dermott, S., Fechtig, H., Graps, A., 
Gustafson, B. A., Hamilton, D. P., Hanner, M. S., Heck, A., 
Hor\'anyi, M., Kissel, J., Lindblad, B.-A., Linkert, D., Linkert, 
G., Mann, I., McDonnell, J. A. M., Morfill, G. E., Polanskey, C., 
Schwehm, G., Srama, R. and Zook, H. A.}
Dust measurements in the jovian magnetosphere. {\em Geophys. Res. Lett.}
{\bf 24}, 2171-2174, 1997

{\bf Gr\"un, E., Kr\"uger, H., Graps, A., Hamilton, D. P., Heck, A.,
Linkert, G., Zook, H. A., Dermott, S., Fechtig, H., Gustafson, B. A.,
Hanner, M. S., Hor\'anyi, M., Kissel, J., Lindblad, B.-A., Linkert, D.,
Mann, I., McDonnell, J. A. M., Morfill, G. E., Polanskey, C., 
Schwehm, G., Srama, R.}  Galileo observes electromagnetically 
coupled dust in the jovian magnetosphere. {\em J. Geophys. Res.}, 
in press, 1998

{\bf Gr\"un, E., Staubach, P., Baguhl, M., Hamilton, D. P., Zook, H. A., 
Dermott, S., Gustafson, B. A., Fechtig, H., Kissel, J., Linkert, D., 
Linkert, G., Srama, R., Hanner, M. S., Polanskey, C., Horanyi, M., 
Lindblad, B.-A., Mann, I., McDonnell, J. A. M., Morfill, G. E. and
Schwehm, G.}
South-north and radial traverses through the zodiacal cloud. {\em Icarus}, 
{\bf 129}, 270-288, 1997c
 
{\bf Hamilton, D.P.\ and Burns, J.A.,} Orbital stability zones about
asteroids~II. The destabilizing effects of eccentric orbits and of
solar radiation. {\em Icarus }{\bf 96}, 43-64, 1992

{\bf Hor\'anyi, M., Gr\"un, E., Heck, A.} Modeling the Galileo
dust measurements at Jupiter. {\em Geophys. Res. Let.} {\bf 24}, 
2175-2178, 1997

{\bf Johnson, T.V., Yeates, C.M.\ and Young, R.,}.
Space Science Reviews Volume on Galileo Mission Overview.
{\em Space Sci.\ Rev. }{\bf 60}, 3-21, 1992

{\bf Kr\"uger, H.,  Gr\"un, E., Landgraf, M., Baguhl, M., Dermott, S., 
Fechtig, H., Gustafson, B. A., Hamilton, D. P., Hanner, M. S., 
Hor\'anyi, M., Kissel, J., Lindblad, B.-A., Linkert, D., Linkert, G.,
Mann, I., McDonnell, J. A. M., Morfill, G. E., Polanskey, C., 
Schwehm, G., Srama, R. and Zook, H. A., }
Three years of Ulysses dust data: 1993 to 1995. 
{\em Planet. Space. Sci.}, 1998, this volume  (Paper~V)

{\bf Landgraf, M. and Gr\"un, E.} In situ measurements of interstellar 
dust. {\em Proceedings of the IAU Colloquium No. 166 on The 
Local Bubble and Beyond}, (edited by D. Breitschwerdt, M.J. Freyberg 
and J. Tr\"umper), Lecture Notes in Physics, Vol. 506, Springer 
Heidelberg, p. 381-384, 1998

{\bf Riemann, R.\ and Gr\"un, E.} Meteor streams, asteroids
and comets near the orbits of Galileo and Ulysses. {\em Proceeding
of the workshop on Hypervelocity Impacts in Space}, (edited by J.A.M.\
McDonnell), University of Kent at Canterbury, 120-125, 1992

{\bf Svestka, J., Auer, S., Baguhl, M. and Gr\"un, E.} 
Measurements of dust electric charges by the Ulysses and Galileo 
dust detectors. In: {\em Physics, Chemistry and Dynamics of Interplanetary 
Dust}, ASP Conference Series, Vol. 104, (edited by B. A. Gustafson and 
M. S. Hanner), page 31-34, 1996

{\bf Zook, H. A., Gr\"un, E., Baguhl, M., Hamilton, D. P., 
Linkert, G., Liou, J.-C., Forsyth, R., Phillips, J. L. }
Solar wind magnetic field bending of jovian dust trajectories. 
{\em Science} {\bf 274}, 1501-1503, 1996.

}

\clearpage

{\small
\begin{table}[htb]
\caption{\label{elements} Heliocentric orbital elements of Galileo's 
interplanetary trajectory for 1993 to 1995. The positional error is 
less than 500,000 km (=~0.003 AU) from 1 Jan 1993 to 1 Jan 1995. The
error increases to  2,500,000 km (=~0.02 AU) by mid 1995
              and to 10,000,000 km (=~0.07 AU) by end of 1995.
The increasing error is due to the strong influence of Jupiter.
}
  \begin{tabular*}{8cm}{ll}
   \hline
   \hline \\[-2.0ex]
Epoch                &    4 Sep. 1993,  16:48:00 \\
Perihelion           &    0.98849~AU   \\
Eccentricity         &    0.68548    \\
Inclination	     &    $ 1.5169^{\circ}$  \\
Long. of Asc. Node   &    $ 255.99^{\circ}$     \\
Arg. of Perihelion   &    $ 186.71^{\circ}$    \\
Mean Anomaly         &    $ 46.953^{\circ}$    \\
True Anomaly         &    $ 130.40^{\circ}$    \\
\hline
\end{tabular*}\\[1.5ex]
\end{table}

\begin{table}[htb]
\caption{\label{event_table} Galileo mission and dust detector (DDS) 
configuration, tests and other events. See text for details. Only 
selected events are given before 1993.
}
{\small
  \begin{tabular*}{15cm}{lccl}
   \hline
   \hline \\[-2.0ex]
Yr-day& 
Date&
Time& 
Event \\[0.7ex]
\hline \\[-2.0ex]
89-291& 18 Oct 1989& 16:52& Galileo launch \\
92-343& 08 Dec 1992& 15:09& Galileo second Earth flyby \\
92-343& 08 Dec 1992& 16:09& DDS configuration: HV=2, EVD=C,I, SSEN=0,0,1,1 \\
92-357& 22 Dec 1992& 14:59& DDS first MRO after second Earth flyby \\
93-166& 15 Jun 1993&      & Galileo 80 bytes MRO Format  \\
93-240& 28 Aug 1993& 16:52& Galileo Ida flyby \\
93-288& 15 Oct 1993& 05:42& DDS noise test start  \\
93-301& 28 Oct 1993& 03:27& DDS noise test end  \\
93-320& 16 Nov 1993& 03:16& DDS noise test start  \\
93-328& 24 Nov 1993& 01:58& DDS noise test end  \\
93-355& 21 Dec 1993& 21:07& DDS noise test start  \\
94-006& 06 Jan 1994& 20:27& DDS noise test end  \\
94-195& 14 Jul 1994& 11:00& DDS last MRO before reprogramming  \\
94-195& 14 Jul 1994& 14:35& DDS counters reset, new event classification program \\
94-196& 15 Jul 1994& 02:00& DDS first MRO after reprogramming  \\
94-197& 16 Jul 1994&      & Galileo start SL~9 observations, duration: 6~days\\
94-210& 29 Jul 1994& 00:14& DDS noise test start  \\
94-219& 07 Aug 1994& 08:14& DDS noise test end  \\
94-300& 27 Oct 1994& 02:38& DDS noise test start  \\
94-312& 08 Nov 1994& 03:08& DDS noise test end  \\
95-019& 19 Jan 1995& 22:56& DDS noise test start  \\
95-029& 29 Jan 1995& 03:09& DDS noise test end  \\
95-035& 04 Feb 1995& 17:44& Galileo phase 1 software load  \\
95-110& 20 Apr 1995& 15:19& DDS noise test start  \\
95-119& 29 Apr 1995& 16:31& DDS noise test end  \\
95-184& 03 Jul 1995& 04:30& DDS last MRO before probe release \\
95-194& 13 Jul 1995&      & Galileo probe release \\
95-205& 25 Jul 1995& 06:54& DDS configuration for ODM wake-up burn: HV=0 \\
95-205& 25 Jul 1995& 13:00& DDS configuration: HV=2  \\  
95-208& 27 Jul 1995& 06:54& DDS configuration for ODM wake-up burn: HV=0 \\
95-208& 27 Jul 1995& 13:00& DDS configuration: HV=2  \\
95-209& 28 Jul 1995& 06:00& DDS first MRO after probe release \\
95-221& 09 Aug 1995& 06:14& DDS noise test start  \\
95-231& 19 Aug 1995& 05:40& DDS noise test end  \\
95-340& 06 Dec 1995& 05:00& DDS last MRO before Io and Jupiter flybys  \\
95-340& 06 Dec 1995& 05:40& DDS configuration: HV=1, EVD=I, SSEN=2,0,2,2  \\
95-341& 07 Dec 1995& 15:21& Galileo start record data (21 bps DDS data)  \\
95-341& 07 Dec 1995& 17:46& Galileo Io flyby, altitude: 892~km  \\
95-341& 07 Dec 1995& 18:25& Galileo end record data  \\
\hline
\end{tabular*}\\[1.5ex]
}
\end{table}

\begin{table}[htb]
{\small
  \begin{tabular*}{15cm}{lccl}
\multicolumn{3}{l}{\normalsize{Table~\ref{event_table} continued.}} & \\
   \hline
   \hline \\[-2.0ex]
Yr-day&
Date&
Time&
Event \\[0.7ex]
\hline \\[-2.0ex]
95-341& 07 Dec 1995& 21:54& Galileo Jupiter closest approach, altitude: 215,000~km  \\
95-341& 07 Dec 1995& 23:22& Galileo start record data (21 bps DDS data)  \\
95-341& 07 Dec 1995& 23:25& DDS configuration: HV=off  \\
95-342& 08 Dec 1995& 00:27& Galileo start orbit insertion burn, duration: 49~min \\
95-342& 08 Dec 1995& 01:26& Galileo end record data  \\
95-362& 28 Dec 1995& 11:00& DDS MRO covering Io and Jupiter flybys  \\[0.7ex]
\hline
\end{tabular*}\\[1.5ex]
}
Abbreviations used: MRO: DDS
memory read-out; HV: channeltron high voltage step; EVD: event definition,
ion- (I), channeltron- (C), or electron-channel (E); SSEN: detection thresholds,
ICP, CCP, ECP and PCP; ODM: orbit deflection maneuver; SL~9: comet Shoemaker-Levy 9.
\end{table}

\begin{table}[htb]
 \begin{center}
\caption[]{\label{class_scheme}Galileo DDS on board classification scheme 
as derived from the analysis of Baguhl et al. (1993). This scheme was implemented 
in the instrument computer onboard Galileo during the 14 July 1994 reprogramming.
Note that 
classes~1 and 2 are now divided into two different subclasses which are mutually 
exclusive. See Paper~I for a detailed explanation of the instrument parameters. 
}
{\scriptsize
  \begin{tabular}{|l||c|cc|cc|c|}
   \hline
   & & & & & &\\
\mc{1}{|l||}{{\bf Parameters}}&
\mc{1}{|c|}{{\bf CLN = 0}} &
\mc{2}{|c|}{{\bf CLN = 1}} &
\mc{2}{|c|}{{\bf CLN = 2}} &
\mc{1}{|c|}{{\bf CLN = 3}} \\
                       &           &           &           &         &   &\\
   \hline
   \hline
   & & & & & &\\
\raisebox{1.2ex}[-1.2ex]{Iongrid}& IA $ > 0 $ & \mc{2}{c|}{IA $ > 0$}  & \mc{2}{c|}{IA $ > 0$} & IA $ > 0$  \\
\raisebox{1.0ex}[-1.0ex]{amplitude\ (IA)}&           &           &          &          &      &     \\
\cline{1-1}                           \cline{3-7}
                       & \raisebox{1.5ex}[-1.5ex]{or}  & \mc{1}{|c|}{}  &  & \mc{1}{|c|}{} & &\\
\raisebox{1.2ex}[-1.2ex]{Target} &  EA $ > 0 $&EA $ > 0 $ & \mc{1}{|c|}{}  & EA $ > 0$ & \mc{1}{|c|}{}& EA $ > 0$ \\
\raisebox{1.0ex}[-1.0ex]{amplitude\ (EA)} &           & \mc{1}{|c|}{}  &    \mc{1}{|c|}{} &     &  \mc{1}{|c|}{}   &  \mc{1}{|c|}{} \\
\cline{1-1}                           \cline{3-7}
                       & \raisebox{1.5ex}[-1.5ex]{or} 
                       & \mc{1}{|c|}{}          & & 
                       & \mc{1}{|c|}{}      & \mc{1}{|c|}{}\\
\raisebox{1.2ex}[-1.2ex]{Channeltron}  & CA $ > 0 $ & \mc{1}{|c|}{} & CA $ > 0 $  &\mc{1}{|c|}{}& CA $ > 0$ & CA $ > 0$  \\
\raisebox{1.0ex}[-1.0ex]{amplitude\ (CA)} &           & \mc{1}{|c|}{} &       &   \mc{1}{|c|}{}    &   &  \\
\hline
                       &           &  & \mc{1}{|c|}{} & \mc{1}{|c|}{} &          & \\
Flighttime target-     &           & \raisebox{1.0ex}[-1.0ex]{EIT\,=\,0} & \mc{1}{|c|}{}& \mc{1}{|c|}{}  &      &\\
iongrid (EIT)          &           & \raisebox{1.5ex}[-1.5ex]{or}  &  \mc{1}{|c|}{}         &
\raisebox{1.5ex}[-1.5ex]{$\rm 3<EIT<15$ }& \mc{1}{|c|}{} &\raisebox{1.5ex}[-1.5ex]{$\rm 3<EIT<15$}\\
                       &           & \raisebox{1.5ex}[-1.5ex]{EIT\,=\,15} &\mc{1}{|c|}{}  & \mc{1}{|c|}{}    &      &  \\
\hline
                       &  & \mc{1}{|c|}{} & & \mc{1}{|c|}{} & &\\
\raisebox{1.2ex}[-1.2ex]{Target-iongrid}         &          & EIC = 1          & \mc{1}{|c|}{}&                   
EIC = 0& \mc{1}{|c|}{} & EIC = 0\\
\raisebox{1.0ex}[-1.0ex]{coincidence (EIC)}     &  &  \mc{1}{|c|}{} &          & \mc{1}{|c|}{}        &  & \\
\hline
                       &  & \mc{1}{|c|}{} &  &\mc{1}{|c|}{} &  &       \\
\raisebox{1.2ex}[-1.2ex]{Channeltron-iongrid}   &     & \mc{1}{|c|}{} &  &   \mc{1}{|c|}{}  &ICC = 1 & ICC = 1\\
\raisebox{1.0ex}[-1.0ex]{coincidence (ICC)}     &           & \mc{1}{|c|}{} &          & \mc{1}{|c|}{} &
 & \\
\hline
                       &           & \mc{1}{|c|}{} &          &   \mc{1}{|c|}{} &        & \\[-1.5ex]
Noise counter of:      &           & \mc{1}{|c|}{} &          &   \mc{1}{|c|}{} &        & \\
\hspace*{0.5cm}target      (EN)  & & \mc{1}{|c|}{}& &EN~$\le $ 8 & \mc{1}{|c|}{} & EN~$ \le $ 8 \\
\hspace*{0.5cm}iongrid     (IN)  & & \mc{1}{|c|}{}& &\hspace{1.8ex}IN~$\le $ 14 &\mc{1}{|c|}{} & \hspace{0.7ex}IN~$\le $ 2 \\
\hspace*{0.5cm}channeltron (CN)  & & \mc{1}{|c|}{}& &\mc{1}{|c|}{} &\hspace{1.1ex}CN~$\le $ 14 &  CN~$ \le $ 2 \\[1.5ex]
\hline
                       &           & \mc{1}{|c|}{}          &   & \mc{1}{|c|}{}        &          & \\[-1.5ex]
EA risetime\ (ET)      &           & \mc{1}{|c|}{}          &    & \mc{1}{|c|}{} &  & $\rm 1\le ET \le 15$ \\[1.5ex]
\hline
                       &           & \mc{1}{|c|}{}          &          &  \mc{1}{|c|}{} &         & \\[-1.5ex]
IA risetime\ (IT)      &           & \mc{1}{|c|}{}          &    & \mc{1}{|c|}{} & & $\rm 1\le IT \le 15$ \\[1.5ex]
\hline
\end{tabular}
}
 \end{center}
\end{table}
\clearpage

\pagestyle{empty}


\begin{sidewaystable}
\tiny
\vbox{
\hspace{-3cm}
\begin{minipage}[t]{22cm}
{\bf Table 4.} Overview of dust impacts accumulated with Galileo DDS 
between 1 January 1993 and 31 December 1995. Day 94-195 when no data were 
received due to the instrument reprogramming is indicated by a horizontal line. 
The heliocentric distance R, 
the lengths of the time interval $\Delta $t (days) from the previous 
table entry, and the corresponding numbers of impacts are given for the 
24 accumulators. The accumulators are arranged with increasing signal 
amplitude ranges (AR), with four event classes for each amplitude 
range (CLN = 0,1,2,3); e.g.~AC31 means counter for AR = 1 and CLN = 3.  
Entries for accumulators that usually contain noise events are marked by 
'$\ast$' signs. Entries marked by '{\footnotesize \#}' denote numbers of impacts 
counted over 25~min intervals. The $\Delta $t in the first line (93-022) 
is the time interval counted from the last entry in Table~3 in 
Paper~II. The totals of  counted impacts, of impacts with 
complete data, and of all events (noise plus impact events) 
for the entire period are given as well. Note that the totals of counted impacts
in amplitude range 1 are not complete due to accumulator overflows during the
dust streams.
\end{minipage} 
}
\bigskip
\hspace{-3cm}
\begin{tabular}{|r|r|r|r|cccc|cccc|cccc|cccc|cccc|cccc|}
\hline
&&&&& &&&&& &&&&& &&&&& &&&&& &&\\
\mc{1}{|c|}{Date}&
\mc{1}{|c|}{Time}&
\mc{1}{|c|}{R}&
\mc{1}{|c|}{$\Delta $t}&
{\scriptsize AC}&{\scriptsize AC}&{\scriptsize AC}&{\scriptsize AC}&
{\scriptsize AC}&{\scriptsize AC}&{\scriptsize AC}&{\scriptsize AC}&
{\scriptsize AC}&{\scriptsize AC}&{\scriptsize AC}&{\scriptsize AC}&
{\scriptsize AC}&{\scriptsize AC}&{\scriptsize AC}&{\scriptsize AC}&
{\scriptsize AC}&{\scriptsize AC}&{\scriptsize AC}&{\scriptsize AC}&
{\scriptsize AC}&{\scriptsize AC}&{\scriptsize AC}&{\scriptsize AC}\\
&&[AU]&\mc{1}{c|}{[d]}&
01&11&21&31&
02&12&22&32&
03&13&23&33&
04&14&24&34&
05&15&25&35&
06&16&26&36\\
&&&&& &&&&& &&&&& &&&&& &&&&& &&\\
\hline
&&&&& &&&&& &&&&& &&&&& &&&&& &&\\
93-022& 16:48&      1.136&25.0&$\ast$&$\ast$&-& 2&$\ast$& 6& 1& 2&-& 3& 1& 4&-& 2&-& 1&-& 2& 1& 2&-& 1&-&-\\
93-039& 22:53&      1.267&17.3&$\ast$&$\ast$&-&-&$\ast$& 3&-& 1&-&-&-& 1&-&-&-& 1&-&-&-&-&-&-&-&-\\
93-061& 02:38&      1.450&21.2&$\ast$&$\ast$&-& 4&$\ast$& 3&-& 3&-& 1&-& 3&-& 3&-& 1&-&-&-& 2&-& 1&-&-\\
\hline
\end{tabular}
\end{sidewaystable}
\clearpage
\begin{sidewaystable}
\tiny
\hspace{-3cm}
\begin{tabular}{|r|r|r|r|cccc|cccc|cccc|cccc|cccc|cccc|}
\hline
&&&&& &&&&& &&&&& &&&&& &&&&& &&\\
\mc{1}{|c|}{Date}&
\mc{1}{|c|}{Time}&
\mc{1}{|c|}{R}&
\mc{1}{|c|}{$\Delta $t}&
{\scriptsize AC}&{\scriptsize AC}&{\scriptsize AC}&{\scriptsize AC}&
{\scriptsize AC}&{\scriptsize AC}&{\scriptsize AC}&{\scriptsize AC}&
{\scriptsize AC}&{\scriptsize AC}&{\scriptsize AC}&{\scriptsize AC}&
{\scriptsize AC}&{\scriptsize AC}&{\scriptsize AC}&{\scriptsize AC}&
{\scriptsize AC}&{\scriptsize AC}&{\scriptsize AC}&{\scriptsize AC}&
{\scriptsize AC}&{\scriptsize AC}&{\scriptsize AC}&{\scriptsize AC}\\
&&[AU]&\mc{1}{c|}{[d]}&
01&11&21&31&
02&12&22&32&
03&13&23&33&
04&14&24&34&
05&15&25&35&
06&16&26&36\\
&&&&& &&&&& &&&&& &&&&& &&&&& &&\\
\hline
&&&&& &&&&& &&&&& &&&&& &&&&& &&\\
94-063& 19:09&      4.045& 8.1&$\ast$&$\ast$&-&-&$\ast$&-&-& 2&-&-&-&-&-&-&-&-&-&-&-&-&-&-&-&-\\
94-074& 13:14&      4.095&10.8&$\ast$&$\ast$&-& 3&$\ast$& 2&-& 1&-&-&-& 1&-&-&-&-&-&-&-&-&-&-&-&-\\
94-085& 13:04&      4.145&11.0&$\ast$&$\ast$&-& 4&$\ast$& 1&-& 1&-&-&-&-&-&-&-&-&-&-&-&-&-&-&-&-\\
\hline
\end{tabular}
\end{sidewaystable}

\clearpage
\begin{sidewaystable}
\tiny
\hspace{-3cm}
\begin{tabular}{|r|r|r|r|cccc|cccc|cccc|cccc|cccc|cccc|}
\hline
&&&&& &&&&& &&&&& &&&&& &&&&& &&\\
\mc{1}{|c|}{Date}&
\mc{1}{|c|}{Time}&
\mc{1}{|c|}{R}&
\mc{1}{|c|}{$\Delta $t}&
{\scriptsize AC}&{\scriptsize AC}&{\scriptsize AC}&{\scriptsize AC}&
{\scriptsize AC}&{\scriptsize AC}&{\scriptsize AC}&{\scriptsize AC}&
{\scriptsize AC}&{\scriptsize AC}&{\scriptsize AC}&{\scriptsize AC}&
{\scriptsize AC}&{\scriptsize AC}&{\scriptsize AC}&{\scriptsize AC}&
{\scriptsize AC}&{\scriptsize AC}&{\scriptsize AC}&{\scriptsize AC}&
{\scriptsize AC}&{\scriptsize AC}&{\scriptsize AC}&{\scriptsize AC}\\
&&[AU]&\mc{1}{c|}{[d]}&
01&11&21&31&
02&12&22&32&
03&13&23&33&
04&14&24&34&
05&15&25&35&
06&16&26&36\\
&&&&& &&&&& &&&&& &&&&& &&&&& &&\\
\hline
&&&&& &&&&& &&&&& &&&&& &&&&& &&\\
95-121& 12:27&      5.223& 7.9&$\ast$&-& 4& 2&$\ast$&-&-&-&$\ast$&-&-& 1&$\ast$&-&-&-&$\ast$&-&-&-&$\ast$&-&-&-\\
95-128& 16:24&      5.231& 7.2&$\ast$&-& 18& 5&$\ast$&-&-&-&$\ast$&-&-& 1&$\ast$&-&-&-&$\ast$&-&-&-&$\ast$&-&-&-\\
95-138& 16:04&      5.241&10.0&$\ast$&-& 8&-&$\ast$&-&-&-&$\ast$&-&-&-&$\ast$&-&-&-&$\ast$&-&-&-&$\ast$&-&-&-\\
\hline
\end{tabular}
\end{sidewaystable}

 \tabcolsep0.07cm                                                               
 \renewcommand{\arraystretch}{0.1}                                              

 \begin{sidewaystable} 
\tiny
\vbox{
\hspace{-4cm}
\begin{minipage}[t]{22cm}
{\bf Table 5.}
 DPF data: No., impact time, CLN, AR, SEC, IA, EA, CA, IT, ET, 
 EIT, EIC, ICC, PA, PET, EVD, ICP, ECP, CCP, PCP, HV and 
 evaluated data: R, LON, LAT, $\rm D_{Jup}$, rotation angle (ROT), instr. pointing 
 ($\rm S_{LON}$, $\rm S_{LAT}$), speed (V), 
 speed error factor (VEF), mass (V) and mass error factor (MEF). For entries with HV = X 
 the channeltron high voltage was switched off.
\end{minipage}
}
\bigskip

\hspace{-4cm}
\begin{tabular*}{26cm}  {@{\extracolsep{\fill}}
rrrrrrrrrrrrrrrcrrrrcrrrrrrrrrrrrrrrrrrrrr}                                     
 \mc{1}{c}{No.}&               \mc{1}{c}{IMP. DATE}&         \mc{1}{c}{C}&                 \mc{1}{c}{AR}&                
 \mc{1}{c}{S}&                 \mc{1}{c}{IA}&                \mc{1}{c}{EA}&                \mc{1}{c}{CA}&                
 \mc{1}{c}{IT}&                \mc{1}{c}{ET}&                \mc{1}{c}{E}&                 \mc{1}{c}{E}&                 
 \mc{1}{c}{I}&                 \mc{1}{c}{PA}&                \mc{1}{c}{P}&                 \mc{1}{c}{E}&                 
 \mc{1}{c}{I}&                 \mc{1}{c}{E}&                 \mc{1}{c}{C}&                 \mc{1}{c}{P}&                 
 \mc{1}{c}{HV}&                \mc{1}{c}{R}&                 \mc{1}{c}{LON}&               \mc{1}{c}{LAT}&               
 \mc{1}{c}{$\rm D_{Jup}$}&     \mc{1}{c}{ROT}&               \mc{1}{c}{$\rm S_{LON}$}&     \mc{1}{c}{$\rm S_{LAT}$}&     
 \mc{1}{c}{V}&                 \mc{1}{c}{VEF}&               \mc{1}{c}{M}&                 \mc{1}{c}{MEF}\\
\end{tabular*} \end{sidewaystable}                                                     
\clearpage                                                                      
\begin{sidewaystable}                                                                   
\tiny \hspace{-4cm}
\begin{tabular*}{26cm}  {@{\extracolsep{\fill}}                                 
rrrrrrrrrrrrrrrcrrrrcrrrrrrrrrrrrrrrrrrrrr}                                     
 \mc{1}{c}{No.}&               \mc{1}{c}{IMP. DATE}&         \mc{1}{c}{C}&                 \mc{1}{c}{AR}&                
 \mc{1}{c}{S}&                 \mc{1}{c}{IA}&                \mc{1}{c}{EA}&                \mc{1}{c}{CA}&                
 \mc{1}{c}{IT}&                \mc{1}{c}{ET}&                \mc{1}{c}{E}&                 \mc{1}{c}{E}&                 
 \mc{1}{c}{I}&                 \mc{1}{c}{PA}&                \mc{1}{c}{P}&                 \mc{1}{c}{E}&                 
 \mc{1}{c}{I}&                 \mc{1}{c}{E}&                 \mc{1}{c}{C}&                 \mc{1}{c}{P}&                 
 \mc{1}{c}{HV}&                \mc{1}{c}{R}&                 \mc{1}{c}{LON}&               \mc{1}{c}{LAT}&               
 \mc{1}{c}{$\rm D_{Jup}$}&     \mc{1}{c}{ROT}&               \mc{1}{c}{$\rm S_{LON}$}&     \mc{1}{c}{$\rm S_{LAT}$}&     
 \mc{1}{c}{V}&                 \mc{1}{c}{VEF}&               \mc{1}{c}{M}&                 \mc{1}{c}{MEF}\\
\end{tabular*} \end{sidewaystable}                                                     
\clearpage                                                                      
\begin{sidewaystable}                                                                   
\tiny \hspace{-4cm}
\begin{tabular*}{26cm}  {@{\extracolsep{\fill}}                                 
rrrrrrrrrrrrrrrcrrrrcrrrrrrrrrrrrrrrrrrrrr}                                     
 \mc{1}{c}{No.}&               \mc{1}{c}{IMP. DATE}&         \mc{1}{c}{C}&                 \mc{1}{c}{AR}&                
 \mc{1}{c}{S}&                 \mc{1}{c}{IA}&                \mc{1}{c}{EA}&                \mc{1}{c}{CA}&                
 \mc{1}{c}{IT}&                \mc{1}{c}{ET}&                \mc{1}{c}{E}&                 \mc{1}{c}{E}&                 
 \mc{1}{c}{I}&                 \mc{1}{c}{PA}&                \mc{1}{c}{P}&                 \mc{1}{c}{E}&                 
 \mc{1}{c}{I}&                 \mc{1}{c}{E}&                 \mc{1}{c}{C}&                 \mc{1}{c}{P}&                 
 \mc{1}{c}{HV}&                \mc{1}{c}{R}&                 \mc{1}{c}{LON}&               \mc{1}{c}{LAT}&               
 \mc{1}{c}{$\rm D_{Jup}$}&     \mc{1}{c}{ROT}&               \mc{1}{c}{$\rm S_{LON}$}&     \mc{1}{c}{$\rm S_{LAT}$}&     
 \mc{1}{c}{V}&                 \mc{1}{c}{VEF}&               \mc{1}{c}{M}&                 \mc{1}{c}{MEF}\\
\end{tabular*} \end{sidewaystable}                                                     
\clearpage                                                                      
\begin{sidewaystable}                                                                   
\tiny \hspace{-4cm}
\begin{tabular*}{26cm}  {@{\extracolsep{\fill}}                                 
rrrrrrrrrrrrrrrcrrrrcrrrrrrrrrrrrrrrrrrrrr}                                     
 \mc{1}{c}{No.}&               \mc{1}{c}{IMP. DATE}&         \mc{1}{c}{C}&                 \mc{1}{c}{AR}&                
 \mc{1}{c}{S}&                 \mc{1}{c}{IA}&                \mc{1}{c}{EA}&                \mc{1}{c}{CA}&                
 \mc{1}{c}{IT}&                \mc{1}{c}{ET}&                \mc{1}{c}{E}&                 \mc{1}{c}{E}&                 
 \mc{1}{c}{I}&                 \mc{1}{c}{PA}&                \mc{1}{c}{P}&                 \mc{1}{c}{E}&                 
 \mc{1}{c}{I}&                 \mc{1}{c}{E}&                 \mc{1}{c}{C}&                 \mc{1}{c}{P}&                 
 \mc{1}{c}{HV}&                \mc{1}{c}{R}&                 \mc{1}{c}{LON}&               \mc{1}{c}{LAT}&               
 \mc{1}{c}{$\rm D_{Jup}$}&     \mc{1}{c}{ROT}&               \mc{1}{c}{$\rm S_{LON}$}&     \mc{1}{c}{$\rm S_{LAT}$}&     
 \mc{1}{c}{V}&                 \mc{1}{c}{VEF}&               \mc{1}{c}{M}&                 \mc{1}{c}{MEF}\\
\end{tabular*} \end{sidewaystable}                                                     
\clearpage                                                                      
\begin{sidewaystable}                                                                   
\tiny \hspace{-4cm}
\begin{tabular*}{26cm}  {@{\extracolsep{\fill}}                                 
rrrrrrrrrrrrrrrcrrrrcrrrrrrrrrrrrrrrrrrrrr}                                     
 \mc{1}{c}{No.}&               \mc{1}{c}{IMP. DATE}&         \mc{1}{c}{C}&                 \mc{1}{c}{AR}&                
 \mc{1}{c}{S}&                 \mc{1}{c}{IA}&                \mc{1}{c}{EA}&                \mc{1}{c}{CA}&                
 \mc{1}{c}{IT}&                \mc{1}{c}{ET}&                \mc{1}{c}{E}&                 \mc{1}{c}{E}&                 
 \mc{1}{c}{I}&                 \mc{1}{c}{PA}&                \mc{1}{c}{P}&                 \mc{1}{c}{E}&                 
 \mc{1}{c}{I}&                 \mc{1}{c}{E}&                 \mc{1}{c}{C}&                 \mc{1}{c}{P}&                 
 \mc{1}{c}{HV}&                \mc{1}{c}{R}&                 \mc{1}{c}{LON}&               \mc{1}{c}{LAT}&               
 \mc{1}{c}{$\rm D_{Jup}$}&     \mc{1}{c}{ROT}&               \mc{1}{c}{$\rm S_{LON}$}&     \mc{1}{c}{$\rm S_{LAT}$}&     
 \mc{1}{c}{V}&                 \mc{1}{c}{VEF}&               \mc{1}{c}{M}&                 \mc{1}{c}{MEF}\\
\end{tabular*} \end{sidewaystable}                                                     
\clearpage                                                                      
\begin{sidewaystable}                                                                   
\tiny \hspace{-4cm}
\begin{tabular*}{26cm}  {@{\extracolsep{\fill}}                                 
rrrrrrrrrrrrrrrcrrrrcrrrrrrrrrrrrrrrrrrrrr}                                     
 \mc{1}{c}{No.}&               \mc{1}{c}{IMP. DATE}&         \mc{1}{c}{C}&                 \mc{1}{c}{AR}&                
 \mc{1}{c}{S}&                 \mc{1}{c}{IA}&                \mc{1}{c}{EA}&                \mc{1}{c}{CA}&                
 \mc{1}{c}{IT}&                \mc{1}{c}{ET}&                \mc{1}{c}{E}&                 \mc{1}{c}{E}&                 
 \mc{1}{c}{I}&                 \mc{1}{c}{PA}&                \mc{1}{c}{P}&                 \mc{1}{c}{E}&                 
 \mc{1}{c}{I}&                 \mc{1}{c}{E}&                 \mc{1}{c}{C}&                 \mc{1}{c}{P}&                 
 \mc{1}{c}{HV}&                \mc{1}{c}{R}&                 \mc{1}{c}{LON}&               \mc{1}{c}{LAT}&               
 \mc{1}{c}{$\rm D_{Jup}$}&     \mc{1}{c}{ROT}&               \mc{1}{c}{$\rm S_{LON}$}&     \mc{1}{c}{$\rm S_{LAT}$}&     
 \mc{1}{c}{V}&                 \mc{1}{c}{VEF}&               \mc{1}{c}{M}&                 \mc{1}{c}{MEF}\\
\end{tabular*} \end{sidewaystable}                                                     
\clearpage                                                                      
\begin{sidewaystable}                                                                   
\tiny \hspace{-4cm}
\begin{tabular*}{26cm}  {@{\extracolsep{\fill}}                                 
rrrrrrrrrrrrrrrcrrrrcrrrrrrrrrrrrrrrrrrrrr}                                     
 \mc{1}{c}{No.}&               \mc{1}{c}{IMP. DATE}&         \mc{1}{c}{C}&                 \mc{1}{c}{AR}&                
 \mc{1}{c}{S}&                 \mc{1}{c}{IA}&                \mc{1}{c}{EA}&                \mc{1}{c}{CA}&                
 \mc{1}{c}{IT}&                \mc{1}{c}{ET}&                \mc{1}{c}{E}&                 \mc{1}{c}{E}&                 
 \mc{1}{c}{I}&                 \mc{1}{c}{PA}&                \mc{1}{c}{P}&                 \mc{1}{c}{E}&                 
 \mc{1}{c}{I}&                 \mc{1}{c}{E}&                 \mc{1}{c}{C}&                 \mc{1}{c}{P}&                 
 \mc{1}{c}{HV}&                \mc{1}{c}{R}&                 \mc{1}{c}{LON}&               \mc{1}{c}{LAT}&               
 \mc{1}{c}{$\rm D_{Jup}$}&     \mc{1}{c}{ROT}&               \mc{1}{c}{$\rm S_{LON}$}&     \mc{1}{c}{$\rm S_{LAT}$}&     
 \mc{1}{c}{V}&                 \mc{1}{c}{VEF}&               \mc{1}{c}{M}&                 \mc{1}{c}{MEF}\\
\end{tabular*} \end{sidewaystable}                                                     
\clearpage                                                                      
\begin{sidewaystable}                                                                   
\tiny \hspace{-4cm}
\begin{tabular*}{26cm}  {@{\extracolsep{\fill}}                                 
rrrrrrrrrrrrrrrcrrrrcrrrrrrrrrrrrrrrrrrrrr}                                     
 \mc{1}{c}{No.}&               \mc{1}{c}{IMP. DATE}&         \mc{1}{c}{C}&                 \mc{1}{c}{AR}&                
 \mc{1}{c}{S}&                 \mc{1}{c}{IA}&                \mc{1}{c}{EA}&                \mc{1}{c}{CA}&                
 \mc{1}{c}{IT}&                \mc{1}{c}{ET}&                \mc{1}{c}{E}&                 \mc{1}{c}{E}&                 
 \mc{1}{c}{I}&                 \mc{1}{c}{PA}&                \mc{1}{c}{P}&                 \mc{1}{c}{E}&                 
 \mc{1}{c}{I}&                 \mc{1}{c}{E}&                 \mc{1}{c}{C}&                 \mc{1}{c}{P}&                 
 \mc{1}{c}{HV}&                \mc{1}{c}{R}&                 \mc{1}{c}{LON}&               \mc{1}{c}{LAT}&               
 \mc{1}{c}{$\rm D_{Jup}$}&     \mc{1}{c}{ROT}&               \mc{1}{c}{$\rm S_{LON}$}&     \mc{1}{c}{$\rm S_{LAT}$}&     
 \mc{1}{c}{V}&                 \mc{1}{c}{VEF}&               \mc{1}{c}{M}&                 \mc{1}{c}{MEF}\\
\end{tabular*} \end{sidewaystable}                                                     
\vfill 

\clearpage


\pagestyle{plain}

\clearpage

\begin{figure}
\epsfxsize=8.5cm
\epsfbox{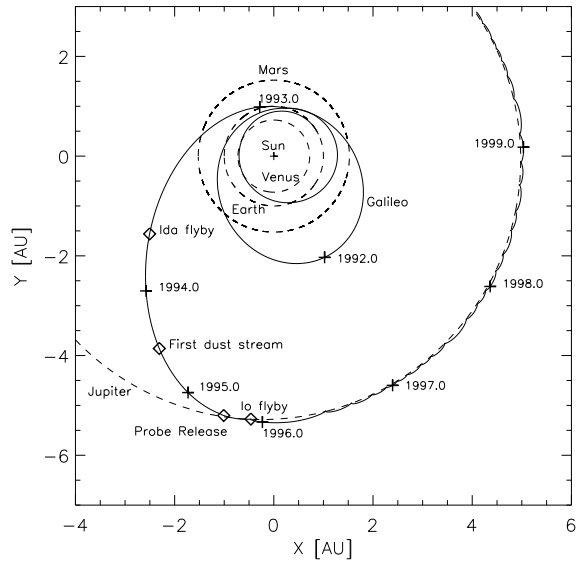}
        \caption{\label{trajectory}
Galileo's interplanetary trajectory from launch until the end of 1999 
(solid line) and the orbits of Venus, Earth, Mars and Jupiter (dashed 
lines). Crosses mark the spacecraft position at the beginning of 
each year, diamonds indicate special events in the time interval 
1993 to 1995 which is the subject of this paper.
}
\end{figure}

\begin{figure}
\epsfxsize=8.5cm
\epsfbox{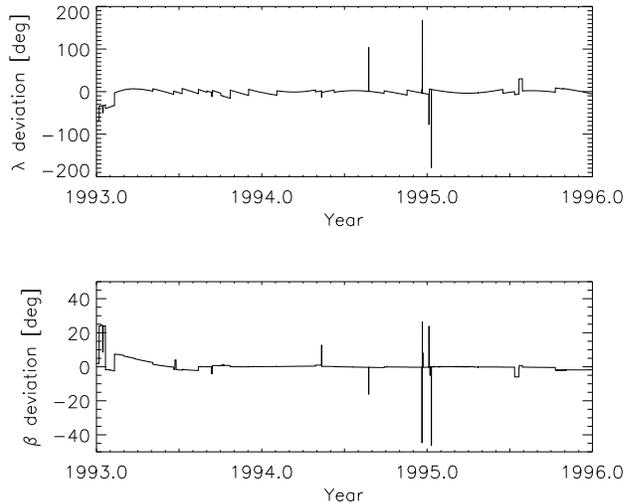}
        \caption{\label{pointing}
Spacecraft attitude: deviation of the antenna pointing direction 
(i.~e. negative spin axis) from the Earth direction. The angles are 
given in ecliptic longitude (top) and latitude (bottom, equinox 1950.0).
}
\end{figure}

\begin{figure}
\epsfxsize=7.5cm
\epsfbox{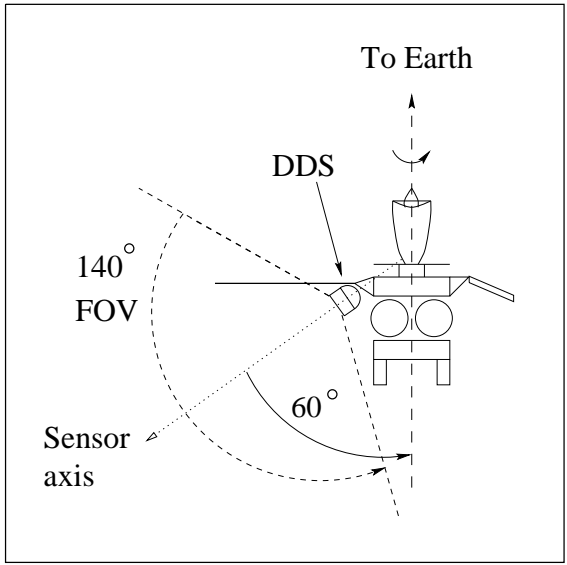}
        \caption{\label{galileo}
Orientation of Galileo and DDS: the antenna (top) points towards Earth and the 
dust detector (DDS) largely faces the anti-Earth hemisphere. The sensor axis has 
an angle of $60^{\circ}$ from the positive spin axis (i.e. 
the anti-Earth direction). 
During one spin revolution
of the spacecraft the sensor axis scans a cone with $120^{\circ}$ 
opening angle. 
The dust detector itself has $140^{\circ}$ field of view 
(FOV). The sensor orientation shown corresponds to a rotation 
angle of $270^{\circ}$ if viewed from the north ecliptic pole.
}
\end{figure}

\begin{figure}
\epsfxsize=8.5cm
\epsfbox{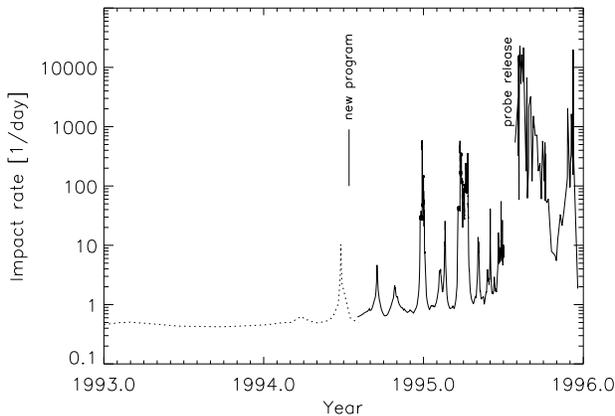}
        \caption{\label{rate}
Total dust impact rate detected by DDS 
as a function of time before  
(dotted line) and after the reprogramming on 14 July 1994 
(solid line). The dotted line is a running average over 9 impacts. 
The data gap in summer 1995 is due to the release of Galileo's 
atmospheric entry probe.
}
\end{figure}

\begin{figure}
\epsfxsize=8.0cm
\epsfbox{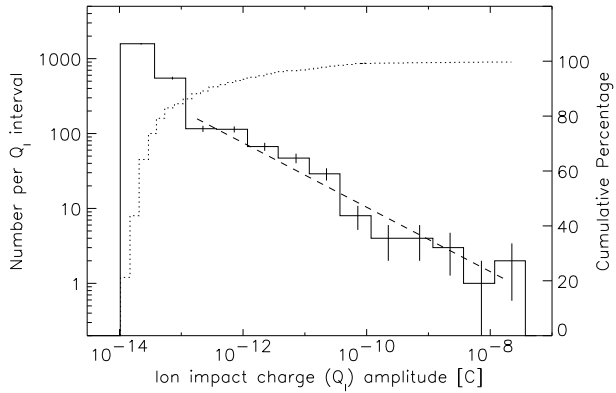}
        \caption{\label{nqi}
Amplitude distribution of the impact charge $\rm Q_I$ for particles 
detected between 1993 and 1995. The solid line
indicates the number of impacts per charge interval, whereas the 
dotted line shows the cumulative distribution. Vertical bars
indicate the $\rm \sqrt{n}$ statistical fluctuation. A power law fit
to the data with $\rm Q_I > 10^{-13}\,C$ is shown as a dashed line 
(power law index -0.43). 
}
\end{figure}

\begin{figure}
\epsfxsize=8.0cm
\epsfbox{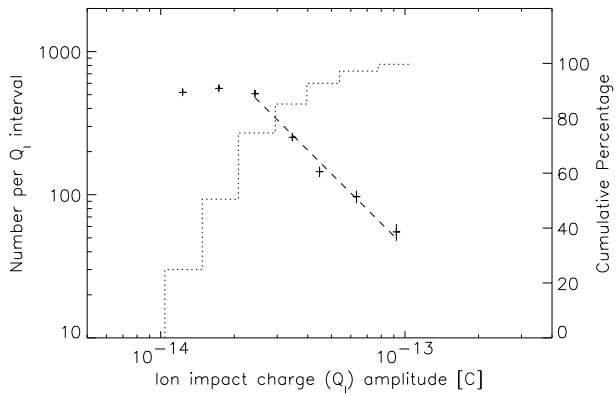}
        \caption{\label{nqi2}
Same as Fig.~\ref{nqi} but for the small particles in the lowest 
amplitude range (AR1) only. A power law fit to the data with 
$\rm 2\times 10^{-14}\,C < Q_I < 10^{-13}\, C$ is shown as a dashed 
line (power law index -1.9).
}
\end{figure}

\begin{figure}
\epsfxsize=8.0cm
\epsfbox{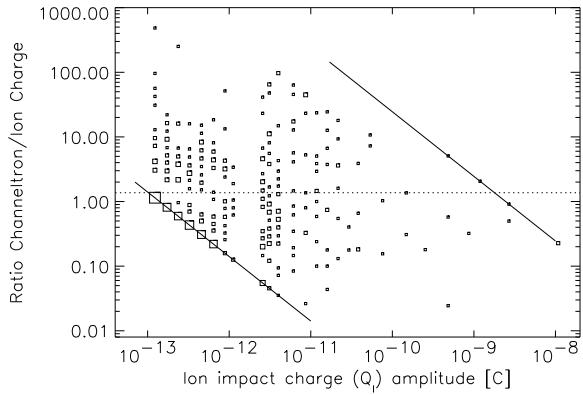}
        \caption{\label{qiqc}
Channeltron amplification factor $\rm A = Q_C/Q_I$
as a function of impact charge $\rm Q_I$ for big particles (AR2 to AR6)
detected between 1993 and 1995. The solid lines indicate the sensitivity
threshold (lower left) and the saturation limit (upper right) of the channeltron. 
Squares indicate dust particle impacts and the area of the squares is proportional 
to the number of events (the scaling of the squares is not the same as in 
in Paper~II). The dotted horizontal line shows the mean value 
of the channeltron amplification A\,=\,1.4 for ion impact charges 
$\rm 10^{-12}~C < Q_I < 10^{-10}~C$.
}
\end{figure}

\begin{figure}
\epsfxsize=8.5cm
\epsfbox{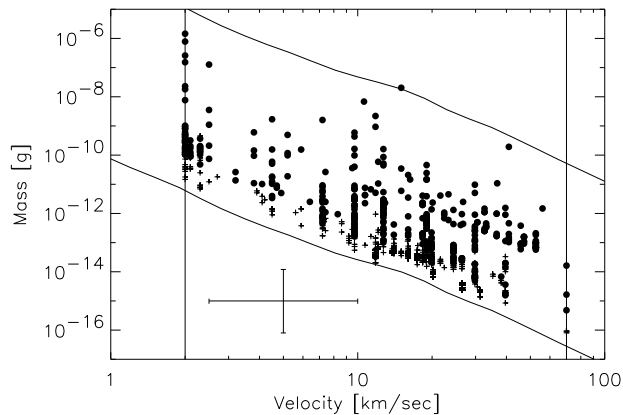}
        \caption{\label{mass_speed}
Masses and impact speeds of all impacts recorded by DDS between 1993 and
1995. The lower and upper solid lines indicate the threshold and
saturation limits of the detector, respectively, and the vertical lines 
indicate the calibrated velocity range. A sample error bar is shown that indicates
a factor of 2 error for the velocity and a factor of 10 for the mass determination.
Note that the small particles (plus signs) are probably faster and smaller 
than implied by this diagram (see text for details).
}
\end{figure}

\begin{figure}
\epsfxsize=8.5cm
\epsfbox{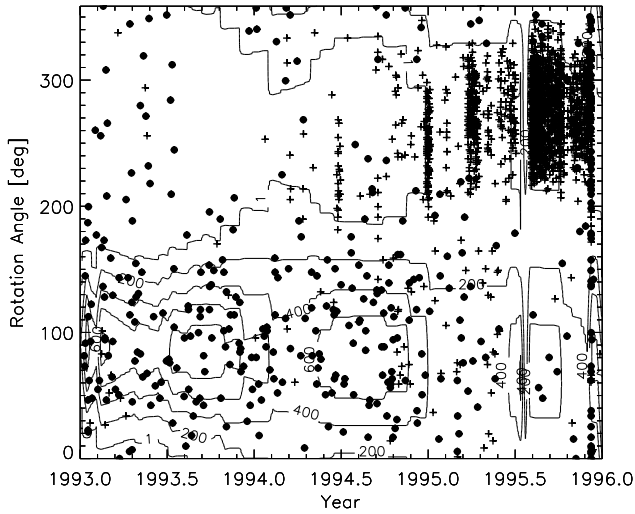}
        \caption{\label{rot_angle} 
Rotation angle vs. time for two different mass ranges (filled circles: 
big particles, AR2 to AR6; plus signs: small particles, AR1). See 
Sect.~\ref{mission} for an explanation of the rotation angle.
The dust streams show up as vertical bands with plus signs.
For some time
periods no rotation angle information was available; these data are
not shown. The contour lines show the sensitive area of DDS for 
interstellar particles (levels of 1, 200, 400, 600 and $\rm 
800\,cm^{2}$ detector area are shown). The number of big 
impacts is depressed during the dust streams because of deadtime 
caused by the large number of small impacts.
}
\end{figure}

\begin{figure}
\epsfxsize=8.5cm
\epsfbox{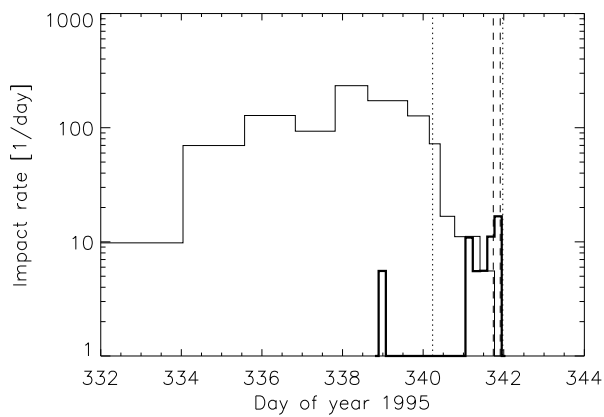}
        \caption{\label{rate_io}
Dust impact rate vs. time for a period of 12 days around Galileo's 
approach towards the inner Jovian system as obtained from the impact 
accumulators. 
The thin solid line shows the impact rate of small particles (AR1)
and the heavy solid line shows that of big particles 
(AR2 to AR6). Because high noise rates in classes 1 and 2 occurred 
on days 340 to 342 only class 3 impacts are shown here. The times 
of closest approaches to Io and Jupiter 
are indicated by dashed lines, and times when the detector 
sensitivity was reduced are shown as dotted lines. Note that the 
peak in the impact rate of big particles on day 339.0 is produced 
by only one impact.
}
\end{figure}

\begin{figure}
\epsfxsize=8.5cm
\epsfbox{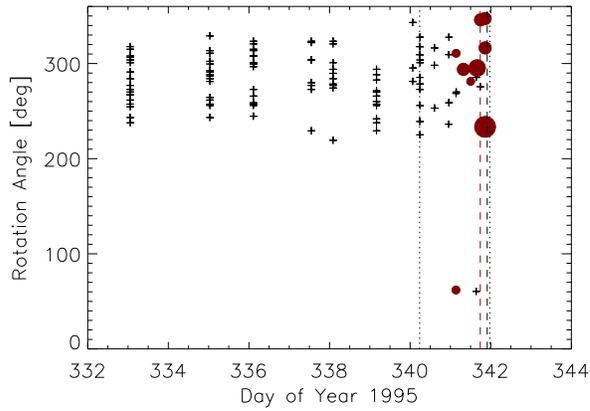}
        \caption{\label{rot_angle_io_1}
Rotation angle vs. time for 12 days around Galileo's 
approach towards the inner Jovian system. Only class~3 impacts 
are shown for which the full information has been transmitted to 
Earth. Plus signs indicate small particles (AR1) 
and filled circles show big particles (the symbol size 
denotes the amplitude range of the particle, AR2 to AR5).  
The times of closest approaches to Io and Jupiter
are indicated by dashed lines, whereas times when the detector
sensitivity was reduced are shown as dotted lines.
The striping before day 341, 15:20~h is due to the
occurrence of MROs once per day which allow for only
4.3~h time resolution and the fact that the instrument
memory of DDS can store only 16 class~3 events. 
Many particles have probably been lost before day 341.
}
\end{figure}

\begin{figure}
\epsfxsize=8.5cm
\epsfbox{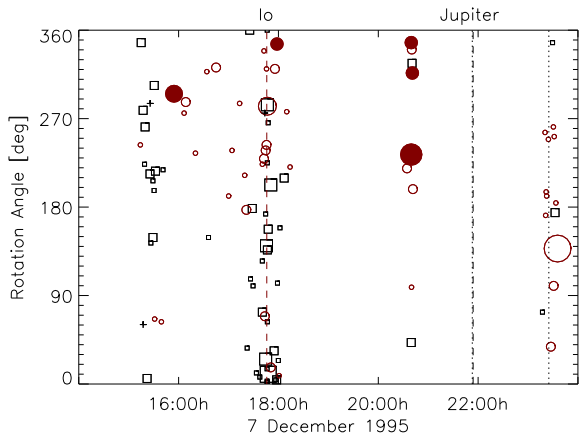}
\caption{
\label{rot_angle_io_2}
Rotation angle vs. time for a period of 12~h around Io closest
approach on 7 Dec 1995 (day 341). The symbols have the following
meaning: ''+'': class~3, AR1; ''{\protect\Large $\protect\bullet$}``:
class~3, AR2 to AR5; ''{\protect\Large $\protect\circ$}'' : class~2;
''$\Box$``: class~1; The size of the symbols indicates the amplitude
ranges of the particles, with AR1 being the smallest and AR6 being the
biggest amplitude range occurring in the diagram. The times of closest
approaches to Io and Jupiter are indicated by dashed lines and the
time when the detector sensitivity was reduced is shown as a dotted
line. Events at 20:40~h occurred in a gap when no data have been
transmitted to Earth and their impact times have 4.3~h
uncertainty. For the other particles the uncertainty in impact time is
usually a few minutes, except for impacts at 15:20~h and 23:20~h which
have 70~min and 33~min uncertainty, respectively (see text for
details).}
\end{figure}

\begin{figure}
\epsfxsize=8.5cm
\epsfbox{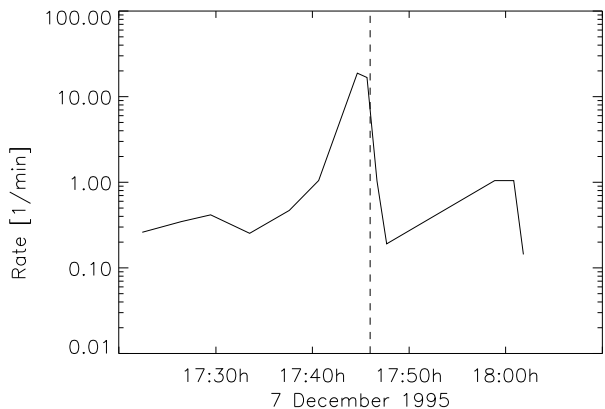}
        \caption{\label{rate_io_2}
Event rate of class~1 events for a period of 50~min around Io closest 
approach. The time of closest approach to Io is indicated by a dashed line.        
} 
\end{figure}

\end{document}